\documentclass[a4paper,aps,pre,showpacs,floatfix,twocolumn,nofootinbib,superscriptaddress]{revtex4-1}
\usepackage{times}

\usepackage[pdftex]{graphicx}
\usepackage[english]{babel}

\usepackage{color}
\usepackage[hidelinks,unicode=true]{hyperref}
\hypersetup{colorlinks=true,linkcolor=blue, urlcolor=blue,
  citecolor=blue,pdfhighlight=/N}

\usepackage{lipsum}

\newcommand{\av}[1]{\langle {#1} \rangle}

\newcommand{\eqref}[1]{(\ref{#1})}

\begin{document}

\title{Random walks in non-Poissoinan activity driven temporal networks}

\author{Antoine Moinet}

\affiliation{Aix Marseille Univ, Universit\'e de Toulon, CNRS, CPT,
  Marseille, France}

\affiliation{Departament de F\'{\i}sica, Universitat Polit\`ecnica de
  Catalunya, Campus Nord B4, 08034 Barcelona, Spain}

\author{Michele Starnini}

\affiliation{ISI Foundation, via Chisola 5, 10126 Torino, Italy}

\author{Romualdo Pastor-Satorras}

\affiliation{Departament de F\'{\i}sica, Universitat Polit\`ecnica de
  Catalunya, Campus Nord B4, 08034 Barcelona, Spain}

\date{\today}

\begin{abstract}
  The interest in non-Markovian dynamics within the complex systems community
  has recently blossomed, due to a new wealth of time-resolved data pointing out
  the bursty dynamics of many natural and human interactions, manifested in an
  inter-event time between consecutive interactions showing a heavy-tailed
  distribution.  In particular, empirical data has shown that the bursty
  dynamics of temporal networks can have deep consequences on the behavior of
  the dynamical processes running on top of them.  Here, we study the case of
  random walks, as a paradigm of diffusive processes, unfolding on temporal
  networks generated by a non-Poissonian activity driven dynamics.  We derive
  analytic expressions for the steady state occupation probability and first
  passage time distribution in the infinite network size and strong aging
  limits, showing that the random walk dynamics on non-Markovian networks are
  fundamentally different from what is observed in Markovian networks.  We found
  a particularly surprising behavior in the limit of diverging average
  inter-event time, in which the random walker feels the network as homogeneous,
  even though the activation probability of nodes is heterogeneously
  distributed.  Our results are supported by extensive numerical simulations.
  We anticipate that our findings may be of interest among the researchers
  studying non-Markovian dynamics of time-evolving complex topologies.
\end{abstract}

\maketitle

\section{Introduction}

Temporal networks~\cite{Holme:2015,lambiotte2016} constitute a recent new
description of complex systems, that, moving apart for the classical static
paradigm of network science~\cite{Newman2010}, in which nodes and edges do not
change in time, consider dynamic connections that can be created, destroyed or
rewired at different time scales.  Within this framework, a first round of
studies proposed temporal network models ruled by homogeneous Markovian
dynamics~\cite{DBLP:journals/corr/abs-1806-04032}.  A prominent example is
represented by the activity-driven model~\cite{Perra:2012}, in which nodes are
characterized by a different degree of \emph{activity}, i.e. the constant rate
at which an agent sends links to other peers, following a Poissonian process.
The memoryless property implied by the Markovian dynamics greatly simplifies the
mathematical treatment of these models, regarding both the topological
properties of the time-integrated network representation
\cite{starnini_topological_2013}, and the description of the dynamical processes
unfolding on activity-driven networks~\cite{perra_random_2012, Starnini2014,
  PhysRevLett.112.118702, Perra_attractiveness, PhysRevE.96.042310,
  Nadini:2018aa}.

However, the Markovian assumption in temporal network modeling has been
challenged by the increasing availability of time-resolved data on different
kinds of interactions, ranging from phone communications~\cite{Onnela:2007} and
face-to-face interactions~\cite{10.1371/journal.pone.0011596}, to natural
phenomena~\cite{PhysRevLett.92.108501, Wheatland1998} and biological
processes~\cite{journals/biosystems/KemuriyamaOSMTKN10}.  These empirical
observations have uncovered a rich variety of dynamical properties, in
particular that the inter-event times $t$ between two successive interactions
(either the creation of the same edge or two consecutive creations of an edge by
the same node), $\psi(t)$, follows heavy-tailed
distributions~\cite{Barabasi:2005,10.1371/journal.pone.0011596,holme2003}.  This
\textit{bursty} dynamics~\cite{Barabasi:2005} is a clear signature that the
homogeneous temporal process description is inadequate and that non-Markovian
dynamics lie at the core of such interactions.  As a consequence, the interest
in non-Markovian dynamics within the complex systems community has recently
blossomed, from the point of view of both mathematical
modeling~\cite{PhysRevLett.114.108701, Karsai:2014, PhysRevE.90.042805,
  Garcia-Perez:2015aa, Jo2014, Kiss:2015aa, Scholtes:2014aa} and dynamical
processes, especially regarding epidemic
spreading~\cite{van_mieghem_non-markovian_2013,PhysRevLett.118.128301,boguna_simulating_2013,Sun:2015}.
Within the framework of non-Markovian networks modeling, the Non-Poissoinan
activity driven (NoPAD) model~\cite{PhysRevLett.114.108701,newnonmarkov} offers
a simple, mathematically tractable framework aimed at reproducing empirically
observed inter-event time distributions, overcoming the limitations of the
classical activity-driven paradigm.

The bursty nature of temporal networks can have a deep impact on dynamical
processes running on top of them, ranging form epidemic spreading, percolation,
social dynamics or synchronization; see Ref.~\cite{porter2016dynamical} for a
bibliographical summary.  Among the many dynamical processes studied on temporal
networks, the random walk stands as one of the most considered, due to its
simplicity and wide range of practical
applications~\cite{WeissRandomWalk,Masuda2017}.  Traditional approaches are
based on the concept of continuous time random walks~\cite{klafter_first_2011},
where the random walk is represented as a renewal process~\cite{renewal}, in which
the probability per unit time that the walker exits a given node through an edge
is constant.  This Poissonian approximation~\cite{klafter_first_2011}, which
translates in a waiting time of the walker inside each node with an exponential
distribution, permits an analytic approach based on a generalized master
equation~\cite{WeissRandomWalk}. The Poissonian case has been considered in
particular for activity-driven
networks~\cite{perra_random_2012,Perra_attractiveness,glassangelica}.

However, if the inter-event time distribution $\psi(t)$ is not exponential, as
empirically observed, the waiting time of the random walker shows aging effects,
meaning that the time at which the walker will leave one node depends on the
exact time at which it arrived at the considered node.  Such memory effects are
particularly important when the inter-event time distribution lacks a first
moment~\cite{aging}. A way to neglect these aging effects is by considering
\textit{active random walks}, in which the inter-event time of a node is
reinitialized when a walker lands on it, in such a way that intervent and
waiting time distributions coincide. In opposition, in \textit{passive random
  walks} the presence of the walker does not reinitialize the inter-event times
of nodes or edges, and thus the waiting time depends on the last activation
time~\cite{2014arXiv1407.4582S}.  The non-Poissoinan scenario has been
considered in the general context of a fixed network in which edges are
established according to a given inter-event time distribution $\psi_{ij}(t)$
for active
walkers~\cite{hoffmann_generalized_2012,2014arXiv1407.4582S,deNigris2016} and
for passive walkers~\cite{2014arXiv1407.4582S}, usually with the assumption of a
finite average inter-event time distribution, with the exception of
Ref.~\cite{deNigris2016}.

In this paper, we contribute to this endeavor with the study of passive random
walks on temporal networks characterized by non-Markovian dynamics, by
considering the case of networks generated by the NoPAD model.  In the NoPAD
model, nodes establish connections to randomly chosen neighbors following a
heavy-tailed inter-event time distribution $\psi_c(t) \sim t^{-1-\alpha}$, with
$0 < \alpha <2$, depending on an activity parameter $c$ assigned to each
node~\cite{PhysRevLett.114.108701,newnonmarkov}.  We show that the dynamics of
passive random walks on NoPAD networks fundamentally departs from the one
observed on classical Poissonian activity-driven networks. For the case
$\alpha >1$, when the average inter-event time is finite, we observe that the
passive random walk behaves in the infinite network limit as an active one with
inter-event time distribution $\psi_c(t) \sim t^{-\alpha}$. For the more
interesting case $\alpha <1$, we argue that a passive random walk behaves, in
the large time limit, as a walker in a homogeneous network. Our results are
checked against extensive numerical simulations.

The paper is organized as it follows: In Section~\ref{sec:random-walks-nopad} we
present the definition of passive random walks on NoPAD
networks. Sect.~\ref{sec:general-formalism} presents a general formalism for the
walker occupation probability and first passage time distribution, that can be
further elaborated in Laplace space in the case $\alpha> 1$, corresponding to an
inter-event time distribution with finite first moment. We present the
application of this formalism for the standard Poissonian activity driven model
in Sec.~\ref{sec:poiss-ad-netw}, recovering previously known results. In
Sec.~\ref{sec:non-poissonian-ad} we consider NoPAD networks with finite average
inter-event times. The case of infinite average inter-event times is discussed
in Sec.~\ref{sec:infin-aver-inter}. Our conclusions are finally presented in
Sec.~\ref{sec:conclusions}.

\section{Passive random walks on NoPAD networks}
\label{sec:random-walks-nopad}

In the NoPAD model ~\cite{PhysRevLett.114.108701,newnonmarkov}, nodes establish
instantaneous connections with randomly chosen peers by following a renewal
process. Each node is activated independently from the others, with the same
functional form of the inter-event time distribution
$\psi_c(t) \sim t^{-\alpha-1}$, with $\alpha>0$, between consecutive activation
events, which depends on an activity parameter $c$, heterogeneously distributed
among the population with a probability distribution $\eta(c)$.  The dynamics of
a random walk on NoPAD networks is defined as follows: A walker arriving at a
node $i$ at time $t$ remains on it until an edge is created joining $i$ and
another randomly chosen node $j$ at a subsequent time $t' > t$, after a waiting time $t' - t$ has
elapsed.  The walker then jumps instantaneously to node $j$ and
waits there until an edge departing from $j$ is created at a subsequent time
$t'' > t'$.  To simplify calculations, here we will focus on \textit{directed
  random walks}: a walker can leave node $i$ only when $i$ becomes active and
creates an edge pointing to another node~\cite{perra_random_2012,glassangelica}.
We consider the case of a passive random walk: the internal clock of the host
node $i$ is not affected by the walker's arrival, and it must wait there until
$i$ creates a new connection.  With this definition, a directed random walk on a
NoPAD network can be mapped to a continuous time random walk on a fully
connected network in which each node has a different distribution of waiting
times.  We assume that all nodes are synchronized at a time $-t_a < 0$ (i.e.,
the internal clock of all nodes is set to zero at time $-t_a$ or, in other
words, we assume the all nodes become active at time $-t_a$) and that the random
walk starts at time $t=0$ from a node with activity $c$, chosen for generality
with probability distribution $H(c)$.

For a general inter-event time distribution $\psi_c(t)$, it is important to
recall that the relevant quantity to describe the motion of a random walker is
the waiting time distribution of residence inside each node. If $\psi_c(t)$
takes an exponential form, the activation rate is constant, implying that the
time to the next activation is independent of the time of the last one. In this
case, the waiting time distribution coincides with $\psi_c(t)$ and memory
effects are absent~\cite{renewal}. When $\psi_c(t)$ has a non-exponential form,
the waiting time distribution is different from $\psi_c(t)$ and indeed it takes
a non-local form: A walker arriving at a node with activity $c$ at time $t$, it
will jump out of it at the next activation event of this node. Assuming the
previous activation event took place at time $t_p < t$, the next one will take
place at time $t_n$, where the inter-event time $\Delta t = t_n - t_p$ is
randomly distributed according to $\psi_c(\Delta t)$. The waiting time of the
walker in node with activity $c$ is thus given by $\tau = t_n -t$ and depends
explicitly on the immediately previous activation time $t_p$. An exact
description of the passive walker will thus require knowledge of the complete
trajectory of the walker in the network, and of the whole sequence of activation
times of all nodes~\cite{2014arXiv1407.4582S}.

This requirement can be however relaxed in the case of NoPAD networks. In the
class of activity driven networks, after an activation event, the walker jumps
to a randomly chosen node. Thus, in the limit of an infinite network, each node
traversed in the path of the walker is essentially visited \textit{for the first
  time}. Therefore, the random walker waiting time distribution depends not on
the whole walker path, but only on the temporal distance to the synchronization
point. In these terms, we consider as the waiting time distribution the
\textit{forward inter-event time} distribution~\cite{klafter_first_2011},
$h_c(t_a+t',t)$, defined as the probability that a walker arriving on a node
with activity $c$ at time $t'$ (hence at a time $t_a+t'$ measured from the
synchronization point of all nodes in the network) will escape from it, due to
the activation of the node, at a time $t'+t$, or, in other words, that it will
wait at the node for a time interval $t$.  When the inter-event time
distribution is exponential, corresponding to a memory-less Poisson process, one
has $h_c(t_a+t',t) \equiv \psi_c(t)$, independent of both the aging time $t_a$
and the arrival time $t'$ of the walker~\cite{renewal}.  For general forms of
the inter-event time distribution $\psi_c(t)$, aging effects take place and the
function $h_c$ depends explicitly on the arrival time $t'$~\cite{aging}.

\section{General formalism}
\label{sec:general-formalism}

In this Section we develop a general formalism to compute the steady state
occupation probability and the first passage time properties of the passive
random walk in infinite NoPAD networks. In the case of an inter-event time
distribution with finite first moment, and in the limit of an infinitely aged
network ($t_a \to \infty$) we can pass to Laplace space to provide closed-form
expressions.

\subsection{Occupation probability}
\label{sec:occup-prob}

We consider here the occupation probability $P(c, t \vert c_0)$, defined as the
probability that a walker is at a node with activity $c$ at time $t$, provided
it started at time $t=0$ on a node with activity $c_0$.  To compute it, let us
define the probability $\Phi_n(c,t \vert c_0)$ that a walker starting at $c_0$
has performed $n$ hops at time $t$, landing at the last hop, made at time $t''$,
with $0 < t'' < t$, at a node with activity $c$.  These two probabilities are
trivially related by the expression
\begin{equation}
  P(c, t \vert c_0) = \sum_{n=0}^\infty \Phi_n(c, t \vert c_0).
  \label{eq:9}
\end{equation}
For $n=0$ (the node $c_0$ does not become activated during the whole time $t$),
we have
\begin{equation}
  \Phi_0(c, t \vert c_0) = \tilde{h}_{c_0}(t_a,t) \delta_{c, c_0},
  \label{eq:11}
\end{equation}
where $\delta_{c, c'}$ is the Kronecker symbol and we have defined
\begin{equation}
  \tilde{h}_{c}(t_a + t', t) = \int_{t}^\infty h_c(t_a + t', \tau) \, d\tau
  \label{eq:12}
\end{equation}
as the probability that a walker arriving at time $t'$ on a node $c$ has not
left it up to time $t+t'$.

To calculate $\Phi_n(c, t \vert c_0)$ for $n\geq 1$ we make use of a
self-consistent condition.  Defining $\Psi_n(t \vert c_0)$ as the probability
that the $n$-th jump of a walker starting at $c_0$ takes place exactly at time
$t$, we can write
\begin{equation}
  \Phi_n(c, t \vert c_0) = \int_0^t \Psi_n(t' \vert c_0)  \eta(c)
  \tilde{h}_{c}(t_a + t',t-t') \, dt'.
  \label{eq:13}
\end{equation}
This equation expresses the sum of the probabilities of the events in which the
walker performs its $n$-th jump at any time $t' < t$, arrives in this jump at a
node $c$, given by the probability $\eta(c)$, and rests at that node $c$ for a
time larger than $t-t'$.  To compute $\Psi_n(t \vert c_0)$ we apply another
self-consistent condition, namely
\begin{equation}
  \Psi_n(t \vert c_0) = \sum_{c'}  \int_0^t \Psi_{n-1}(t' \vert c_0) \eta(c')
  h_{c'}(t_a+t',t-t') \, dt'.
  \label{eq:15}
\end{equation}
This equation implies the $(n-1)$-th jump taking place at time $t'$, and landing
on a node $c'$, with probability $\eta(c')$, and the last jump taking place,
from $c'$, at time $t-t'$.  The expression is averaged over all possible values
of the activity $c'$ of the intermediate step. The iterative Eq.~(\ref{eq:15}),
complemented with the initial condition
$\Psi_1(t \vert c_0) = \tilde{h}_{c_0}(t_a,t)$, provides a complete solution for
the steady state probability, via Eqs.~(\ref{eq:13}) and~(\ref{eq:9}).

\subsection{First passage time distribution}
\label{sec:first-passage-time}

We now consider the first passage time probability $F(t, c\vert c_0)$, defined
as the probability that a walker starting at a node of activity $c_0$ arrives
\textit{for the first time} at another node of activity $c$ exactly at time
$t$. To compute it, we define $\bar{\Psi}_n(t |c, c_0)$ as the probability that
the walker performs his $n$-th hop at time $t$, irrespective of where it lands,
in a trajectory that has never visited before a node of activity $c$. We can
thus write\footnote{We neglect here the case $c_0 = c$. Its consideration will
  imply an additional term $\delta(t) \delta_{c_0, c}$ in Eq.~(\ref{eq:8}),
  where $\delta(t)$ is the Dirac delta function.}
\begin{widetext}
\begin{equation}
\label{eq:8}
  F(t, c\vert c_0) = h_{c_0}(t_a,t)\eta(c) + \int_0^{t} dt' \sum_{n=1}^\infty
                       \bar{\Psi}_n(t' |c, c_0)
  \sum_{c' \neq c} \eta(c') h_{c'}(t_a + t' , t - t')\eta(c).
\end{equation}
\end{widetext}
In this equation, the first term accounts for the walker arriving at $c$ in a
single hop, while for the second term we consider that the walker has performed
an arbitrary number of hops $n\geq 1$ at a time $t'$, that the last of these
hops lands on a node with activity $c' \neq c$, and from this node the walker
performs a final hop, after a waiting time $t-t'$ that lands it on a node with
activity $c$.  The probability $\bar{\Psi}_n(t | c,c_0)$ can be recovered from
the recurrent relation
\begin{equation}
  \label{eq:7}
  \bar{\Psi}_n(t | c,c_0) =  \int_0^{t} dt'  \bar{\Psi}_{n-1}(t' | c,c_0)
  \sum_{c' \neq c}   \eta(c')  h_{c'}(t_a + t' , t - t'),
\end{equation}
which considers the $(n-1)$-th hop taking place at time $t'$, landing at a node
of activity $c' \neq c$, and performing a last hop after a time $t-t'$.

\subsection{Inter-event time distributions with finite average}
\label{sec:finite-average-inter-1}

While the previous formalism is exact for NoPAD networks of infinite size, it
cannot be developed further in absence of detailed information about the
functional form of the forward inter-event time distribution $h_c(t', t)$, which
is in general very hard to obtain~\cite{klafter_first_2011}. Progress is
possible, however, when the first moment of the inter-event time distribution
$\psi_c(t)$, defined as
\begin{equation}
  \label{eq:5}
  \bar{\tau_c} = \int_0^\infty u\,\psi_c(u) \;du,
\end{equation}
is finite.  When this condition applies, and in the limit of very large aging
time $t_a \to \infty$, the forward inter-event time distribution does no longer
depend on its first argument and it is given
by~\cite{renewal,klafter_first_2011,burstylambiotte2013}
\begin{equation}
  \label{eq:32}
  h_c (t)  = \frac{1}{\bar{\tau}_c}\int_{t}^{\infty}\psi_c (u)du.
\end{equation}
In this double limit of infinite network size and aging time, the passive random
walker behaves effectively as an active random walker in which the waiting time
distribution is given the forward inter-event time distribution $ h_c (t)$.

Under assumption of a finite average inter-event time, defining the Laplace
transforms
\begin{eqnarray}
  \label{eq:4}
  \Phi_n(c, s \vert c_0) &=& \int_0^\infty \Phi_n(c, t \vert c_0) e^{-st}
                          dt,\\
  \Psi_n(s \vert c_0) &=& \int_0^\infty  \Psi_n(t \vert c_0)  e^{-st}
                          dt,\\
  h_{c}(s) &=&  \int_0^\infty   h_{c}(t)  e^{-st},
  \\
  \tilde{h}_{c}(s) &=&  \int_0^\infty   \tilde{h}_{c}(t)  e^{-st}
                       dt,
\end{eqnarray}
we can write Eq.~\eqref{eq:13} in Laplace space as
\begin{equation}
  \Phi_n(c, s \vert c_0) = \eta(c) \Psi_n(s \vert c_0) \tilde{h}_{c}(s),
  \label{eq:14}
\end{equation}
while Eq.~\eqref{eq:15} takes the form
\begin{equation}
  \Psi_n(s \vert c_0) = \sum_{c'} \eta(c') \Psi_{n-1}(s  \vert
  c_0)h_{c'}(s).
  \label{eq:16}
\end{equation}
Eq.~(\ref{eq:16}) can be easily solved, yielding
\begin{equation}
  \Psi_n(s \vert c_0) = \left[ \sum_{c'} \eta(c') h_{c'}(s) \right]^{n-1}
    \Psi_1(s \vert c_0).
    \label{eq:17}
\end{equation}
Considering that $\Psi_1(t \vert c_0) = h_{c_0}(t)$, we can combine
Eqs.~(\ref{eq:17}) and~(\ref{eq:14}) to obtain
\begin{eqnarray}
  P(c, s \vert c_0)
  & =&  \sum_{n=0}^\infty \Phi_n(c, s \vert c_0)   =
       \tilde{h}_{c_0}(s) \delta_{c, c_0} \nonumber \\
  &+& \eta(c) \tilde{h}_{c}(s)  h_{c_0}(s)
      \sum_{n=1}^\infty   \left[ \sum_{c'}  \eta(c') h_{c'}(s)
      \right]^{n-1}   \nonumber \\
  &=& \tilde{h}_{c_0}(s) \delta_{c, c_0} +  \frac{\eta(c)  \tilde{h}_{c}(s)
      h_{c_0}(s)}{1 - \sum_{c'} \eta(c')
      \tilde{h}_{c'}(s)}. \label{eq:6}
\end{eqnarray}
From Eq.~(\ref{eq:6}) we can obtain the probability $P(c, t)$ of observing the
walker at a node $c$ at time $t$, irrespective of the position $c_0$ of origin,
as
\begin{equation}
  P(c, t) = \sum_{c_0} H(c_0) P(c, t \vert c_0),
  \label{eq:10}
\end{equation}
which, from Eq.~(\ref{eq:6}), can be written in Laplace space as
\begin{equation}
   P(c, s) = H(c) \tilde{h}_{c}(s) + \frac{\eta(c) \tilde{h}_{c}(s)
     \sum_{c_0}H(c_0) h_{c_0}(s)}{1 -
     \sum_{c'}\eta(c')h_{c'}(s)}.  \label{eq:c}
\end{equation}

For the first passage time distribution, Laplace transforming Eq.~(\ref{eq:7}),
we obtain
\begin{equation}
  \label{eq:18}
  \bar{\Psi}_n(s | c,c_0) = \bar{\Psi}_{n-1}(s | c,c_0)  \sum_{c' \neq c}
  \eta(c')  h_{c'}(s).
\end{equation}
Since $ \bar{\Psi}_1(s | c,c_0) = h_{c_0}(s)$, we have, solving the
recursion relation Eq.~(\ref{eq:18}),
\begin{equation}
  \label{eq:19}
   \bar{\Psi}_n(s | c,c_0) =  h_{c_0}(s) \left(  \sum_{c' \neq c}
     \eta(c') h_{c'}(s) \right)^{n-1}.
\end{equation}
Introducing Eq.~(\ref{eq:19}) into the Laplace space counterpart of
Eq.~(\ref{eq:8}), we finally have
\begin{widetext}
\begin{equation}
\label{eq:20}
  F(s,c\vert c_0) =  h_{c_0}(s)\eta(c)+ h_{c_0}(s) \eta(c) \frac{\sum_{c' \neq c}
    \eta(c') h_{c'}(s)}{1- \sum_{c' \neq c}
    \eta(c') h_{c'}(s)}
    = \frac{h_{c_0}(s)\eta(c)}{1- \sum_{c' \neq c}
    \eta(c') h_{c'}(s)}
\end{equation}
\end{widetext}
The mean first passage time (MFPT), defined as
\begin{equation}
  \label{eq:21}
  T(c\vert c_0) = \int_0^\infty t\,F(t,c\vert c_0) dt,
\end{equation}
can be obtained, from the Laplace transform in Eq.~(\ref{eq:20}),
as~\cite{klafter_first_2011}
\begin{equation}
  \label{eq:22}
  T(c\vert c_0) = - \left. \frac{ d F(s, c\vert c_0)}{d s} \right|_{s = 0}.
\end{equation}

\section{Poissonian Activity-Driven networks}
\label{sec:poiss-ad-netw}

To check the expressions obtained in the previous Section, we start by
considering Poissonian AD networks, for which the random walk problem has been
already studied~\cite{perra_random_2012,glassangelica}.  In this case, the
inter-event time distributions for each node takes an exponential form,
$\psi_c(t) = c e^{-c t}$, where the value of $c$ is extracted from the
distribution $\eta(c)$.  This fact renders the results in
Section~\ref{sec:general-formalism} exact.  The system completely lacks memory, and
it holds $h_c(t', t) \equiv \psi_c(t) = c e^{-c t}$, while
$\tilde{h}_c(t' | t) = e^{-ct}$. With the corresponding Laplace transforms
$h_c(s) = c/(c+s)$ and $\tilde{h}_c(s) = 1/(c+s)$, the probability $P(c, t)$ of
observing the walker in a node $c$ at time $t$ can be written, from
Eq.~(\ref{eq:c}), as
\begin{equation}
  \label{eq:23}
  P(c, s) = \frac{H(c)}{c+s} + \frac{\eta(c)}{c+s} \frac{\sum_{c_0}
    H(c_0) \frac{c_0}{c_0+s}}{1 - \sum_{c'} \eta(c') \frac{c'}{c'+s} }.
\end{equation}
The steady state occupation probability
$P_\infty(c) = \lim_{t \to \infty} P(c, t)$ can be obtained in Laplace space as
the alternative limit $P_\infty(c) = \lim_{s \to 0} s P(c, s)$, leading to
\begin{equation}
  \label{eq:25}
  P_\infty(c) = \frac{1}{\av{c^{-1}}}\frac{\eta(c)}{c},
\end{equation}
independent of the initial distribution $H(c_0)$, where
$\av{c^{-1}} = \sum_{c'} \eta(c')/c'$ is the average of the inverse activity in
the network.  Thus, as time increases, the average occupation probability
crosses over from the initial distribution at time $t=0$ of random walkers,
$P_0(c) = H(c)$, to the steady state occupation probability,
$P_\infty(c) \sim \eta(c) / c$, at large times~\cite{glassangelica}.

The first passage time probability in Laplace space, Eq.~(\ref{eq:20}), in the
Poissonian case, reads
\begin{equation}
  \label{eq:26}
  F(s,c\vert c_0) = \frac{c_0\,\eta(c)}{(c_0 + s)\left(1 - \sum_{c'
        \neq c} \eta(c') \frac{c'}{c'+s} \right)}.
\end{equation}
The first derivative of Eq.~(\ref{eq:26}) evaluated at $s=0$ leads to the MFPT
on a node with activity $c$, when the walker starts from a node with activity
$c_0$,
\begin{eqnarray}
  \label{eq:27}
  T(c\vert c_0) =\frac{1}{\eta(c)}\left[\av{c^{-1}}-\frac{\eta(c)}{c}\right] +
  \frac{1}{c_0}.
\end{eqnarray}
The MFPT of nodes with activity $c$, irrespective of the activity $c_0$ of the
starting node, can be obtained by averaging over the initial position of the
walker, $T(c) = \sum_{c_0}H(c_0) T(c\vert c_0)$.  If such position is chosen
uniformly at random in the network, $H(c_0) = \eta(c_0)$, Eq.~(\ref{eq:27})
becomes
\begin{equation}
\label{eq:27b}
 T(c) =  \frac{\av{c^{-1}}}{\eta(c)}  -\frac{1}{c} + \av{c^{-1}}.
\end{equation}
This expression provides a correction to the result in
Ref.~\cite{perra_random_2012}, derived by a pure mean-field calculation.  The
first term of Eq.~(\ref{eq:27b}) can be obtained by following the mean field
argument in Ref.~\cite{perra_random_2012}, and it indicates that the MFPT of
nodes with activity $c$ will be inversely proportional to the density of nodes
in that activity class $c$, given by $\eta(c)$. The second term takes into
account the probability of not arriving earlier on nodes with activity $c$,
while the third term, which is constant in $c$, accounts for the escape time
from the starting node of the walker.  We note that, while the second term is
always negligible with respect to the others, the third constant term can be
relevant for nodes of small activity $c$, if the activity distribution $\eta(c)$
is power law distributed, $\eta(c) \sim c^{-\gamma}$~\cite{Perra:2012}.

\subsection{Numerical Application}
\label{sec:application}

To provide an example application, we consider the simplest case of an AD
network with two different activities $1$ and $\epsilon < 1$, with an activity
distribution
\begin{equation}
\eta(c) = p \,\delta_{c, \epsilon} + (1-p)\,\delta_{c, 1}.
\label{eq:bi-valued}
\end{equation}
If we assume $H(c) = \eta(c)$, Eq.~(\ref{eq:23}) reduces to
\begin{equation}
  \label{eq:24}
  P(c, s) =  \frac{\eta(c)}{s(c+s)} \left( \sum_{c'}
    \frac{\eta(c')}{c'+s} \right)^{-1}.
\end{equation}
From here, we can obtain
\begin{equation}
  \label{eq:1}
  \sum_{c'} \frac{\eta(c')}{c' + s} = \frac{(1-p)\epsilon + p +
    s}{(\epsilon+s)(1+s)},
\end{equation}
which leads to
\begin{eqnarray}
  \label{eq:2}
  P(\epsilon, s) &=& \frac{p(1+s)}{s [(1-p)\epsilon + p + s]} \nonumber \\
  &\equiv&   p t_c \frac{1}{s} - t_c p(1-p)(1-\epsilon)\frac{1}{t_c^{-1} + s},
\end{eqnarray}
with $t_c = p + (1-p)\epsilon$.  This expression in Laplace space can be
trivially anti-transformed, yielding the occupation probability
\begin{equation}
  \label{eq:3}
  P(\epsilon, t) = P_\infty(\epsilon) - t_c p(1-p)(1-\epsilon) e^{-t/t_c},
\end{equation}
with
$P_\infty(\epsilon) = \eta(\epsilon)/[\epsilon\av{c^{-1}}] = p/[(1-p)\epsilon +
p]$.  The occupation probability relaxes exponentially to the steady state with
a time scale $t_c$ that can become very small when both $p$ and $\epsilon$ tend
to zero~\cite{glassangelica}.

For the first passage time distribution, application of Eq.~(\ref{eq:26}) leads
directly to
\begin{equation}
  \label{eq:29}
  F(s, \epsilon | 1) = \frac{p}{p+s},
\end{equation}
indicating an exponential distribution in real time
$F(t, \epsilon| 1) = p e^{-pt}$. The MFPT is obtained as
\begin{equation}
  \label{eq:28}
  T(\epsilon | 1) = \frac{1}{p},
\end{equation}
independent of $\epsilon$.  On the other hand,
\begin{equation}
  \label{eq:30}
  F(s, 1 | \epsilon) = \frac{\epsilon(1-p)}{\epsilon(1-p) + s},
\end{equation}
leading to $F(t, 1 | \epsilon) = \epsilon(1-p)e^{-\epsilon(1-p) t}$, from which
one can obtain the MFPT
\begin{equation}
  \label{eq:31}
  T(1 | \epsilon) = \frac{1}{\epsilon(1-p)},
\end{equation}
diverging in the limits $\epsilon \to 0$ or $p \to 1$.

\section{Non-Poissonian Activity-Driven networks with finite average inter-event
  time}
\label{sec:non-poissonian-ad}

We now consider the more interesting case of non-Poissonian Activity-Driven
(NoPAD) networks, in which the inter-event time distribution is different from
exponential.  To fix  notation, we will focus in particular in the power law
form
\begin{equation}
  \psi_c (t) = \alpha c (c t +1)^{-(1+\alpha)},
\label{eq:d}
\end{equation}
with $\alpha > 0$ to allow for normalization. Here we will consider the case
$\alpha > 1$ corresponding to finite  average inter-event time of value
\begin{equation}
  \label{eq:34}
  \bar{\tau_c} = \int_0^\infty u\,\psi_c(u) \;du  = (\alpha-1) c (c t +1)^{-\alpha}.
\end{equation}

In this case, for an infinitely aged network, $t_a\rightarrow +\infty$, the
forward recurrence time no longer depends on the aging time and one has, from
Eq.~(\ref{eq:32}),
\begin{eqnarray}
 h_c (t) & = \frac{1}{\bar{\tau}_c}\int_{t}^{\infty}\psi_c (u)du =
 (\alpha-1)c\,(ct+1)^{-\alpha}, \label{eq:33} \\
   \tilde{h}_{c}(t) & = \int_{t}^{\infty}h_c (u)du = (1+ct)^{1-\alpha}.
\end{eqnarray}
In the limit of large $t \gg 1$, which correspond to $s \ll 1$ in the Laplace
space, and by virtue of the Tauberian theorems~\cite{klafter_first_2011}, we can
write, for $1 < \alpha < 2$,
\begin{eqnarray}
  \psi_c(s)
  &\simeq &  1-\bar{\tau}_c\, s +\frac{\Gamma_{2-\alpha}}{\alpha-1}
            \left(\frac{s}{c}  \right)^\alpha  +
            o(s^\alpha)  \label{eq:laplace_psi},\\
  h_c (s)
  & \simeq &
             1-\Gamma_{2-\alpha}  \left(  \frac{s}{c}
             \right)^{\alpha-1} + o(s^{\alpha-1}),\\
  \tilde{h}_{c}(s) &\simeq&
                            \frac{\Gamma_{2-\alpha}}{c} \left(  \frac{s}{c}
                            \right)^{\alpha - 2} + o(s^{\alpha-2}) ,
\end{eqnarray}
where $o(x)$ denotes a function $f(x)$, such that
$\lim_{x\to0} \frac{f(x)}{x} = 0$, and $\Gamma_z \equiv \Gamma(z)$ is the Gamma
function \cite{abramovitz}.  These expressions, combined with Eq.~\eqref{eq:c},
yield
\begin{equation}
  P(c,s) \simeq
  \frac{\eta(c)c^{-(\alpha-1)}}{s\av{c^{-(\alpha-1)}}}+o\left( s^{-1} \right) .
\end{equation}
By taking the limit $t \to \infty$, the steady state occupation probability
finally reads
\begin{equation}
  P_\infty(c) = \frac{\eta(c)c^{-(\alpha-1)}}{\av{c^{-(\alpha-1)}}}.
\label{eq:steady_state1}
\end{equation}

\begin{figure}[t]
\centering
  \includegraphics[width=\columnwidth]{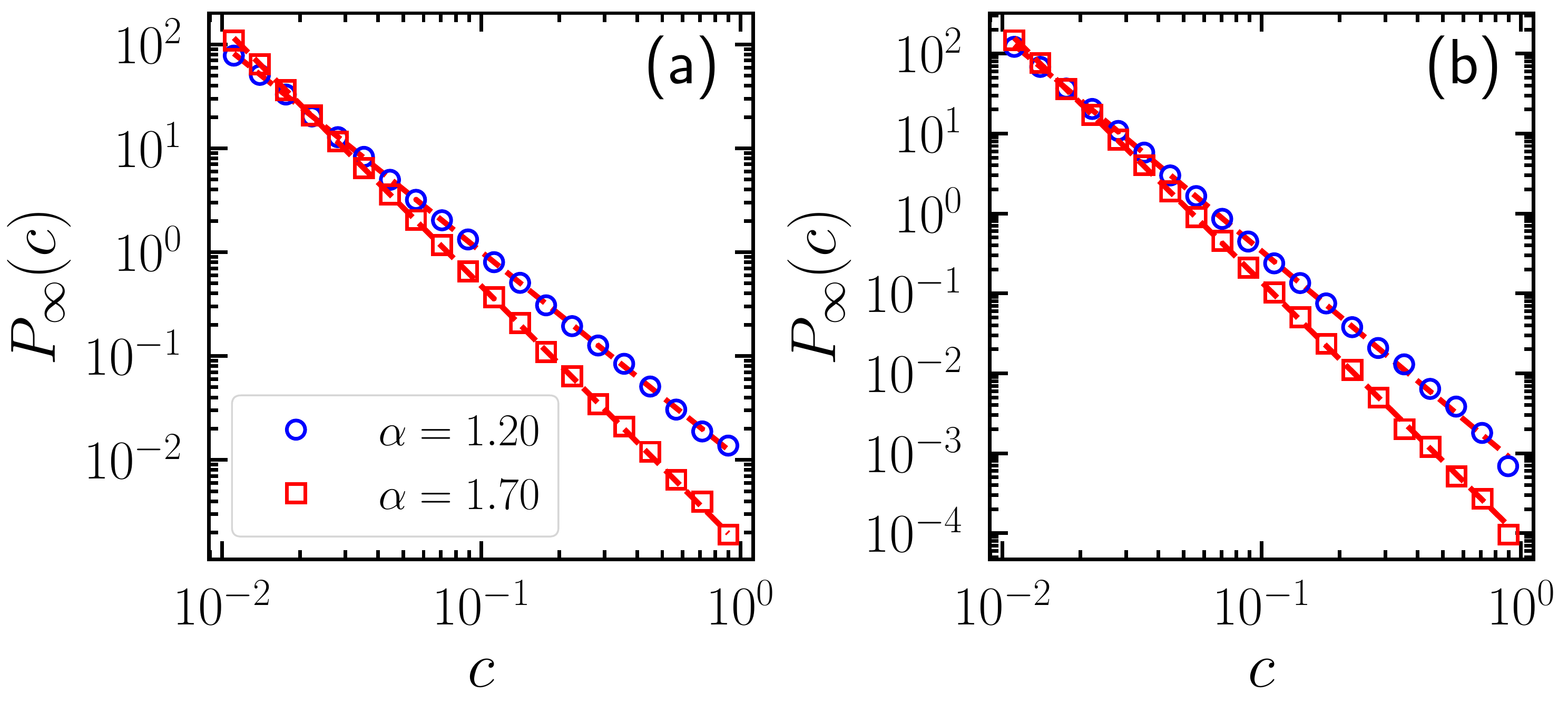}
  \caption{Steady state occupation probability of a directed passive random walk
    on an infinitely aged NoPAD network, $P_\infty(c)$, with inter-event time distribution
    given by Eq.~\eqref{eq:d}, for different values of the exponent $\alpha$. We
    consider an activity distribution $\eta(c) \sim c^{-\gamma}$ with
    $\epsilon=10^{-2}$ and exponent (a) $\gamma = 1.80$ and (b) $\gamma=2.50$.
    Dashed lines correspond to the analytical result of
    Eq.~\eqref{eq:steady_state1}. The steady state distribution is computed from
    $W = 10^6$ different walks, stopped at time $t=10^6$. Network size
    $N=10^5$}.
  \label{fig:steadystate}
\end{figure}

In order to confirm this result, we have performed numerical simulations of the
passive directed random walk on an NoPAD network with an inter-event time
distribution given by Eq.~\eqref{eq:d} and power-law distributed activity
$\eta(c)\propto c^{-\gamma}$, where the activity takes values in the interval
$[\epsilon,1]$. In Fig.~\ref{fig:steadystate} we show the theoretical prediction
Eq.~(\ref{eq:steady_state1}) (dashed lines), obtained in the limit of infinite
network size, compared with direct simulation in finite networks of size
$N=10^{5}$ for different values of the inter-event time exponent $\alpha > 1$
(hollow symbols). Simulations are performed in the limit of infinite aging time,
$t_a \to \infty$. The first activation of every node takes thus place at a time,
measured from the beginning of the random walk $t=0$, given by the distribution
in Eq.~(\ref{eq:33}). We keep track of the whole history of the network, each
successive activation of each node taking place at inter-event times given by
Eq.~(\ref{eq:d}). Walks are stopped at time time $t=10^6$, where the occupation
probability is computed. As we can see, even for the moderate network size
considered here, the infinite network approximation provides an excellent
approximation for the steady state distribution.

Interestingly, when taking the limit $\alpha \rightarrow 2$ in
Eq.~\eqref{eq:steady_state1}, we recover the result established for Poissonian
AD networks, i.e. $P_\infty(c)=\eta(c)c^{-1}/\av{c^{-1}}$.  This result is
general for any $\alpha \geq 2$, as can be seen from the corresponding leading
order expansions in Laplace space for this range of $\alpha$ values, namely,
\begin{eqnarray}
  \psi_c(s) &\simeq
  & 1-\bar{\tau}_c\, s + \frac{1}{2} \bar{\tau_c^2} \, s^2, \\
  h_c (s) & \simeq &
              1- \frac{1}{c(\alpha -2)} \, s, \label{eq:36}\\
\tilde{h}_{c}(s) & \simeq & \frac{1}{c(\alpha -2)},
\end{eqnarray}
which, substituted on Eq.~\eqref{eq:c}, lead again to Eq.~\eqref{eq:25} in the
steady state.

\begin{figure}[t]
\centering
  \includegraphics[width=\columnwidth]{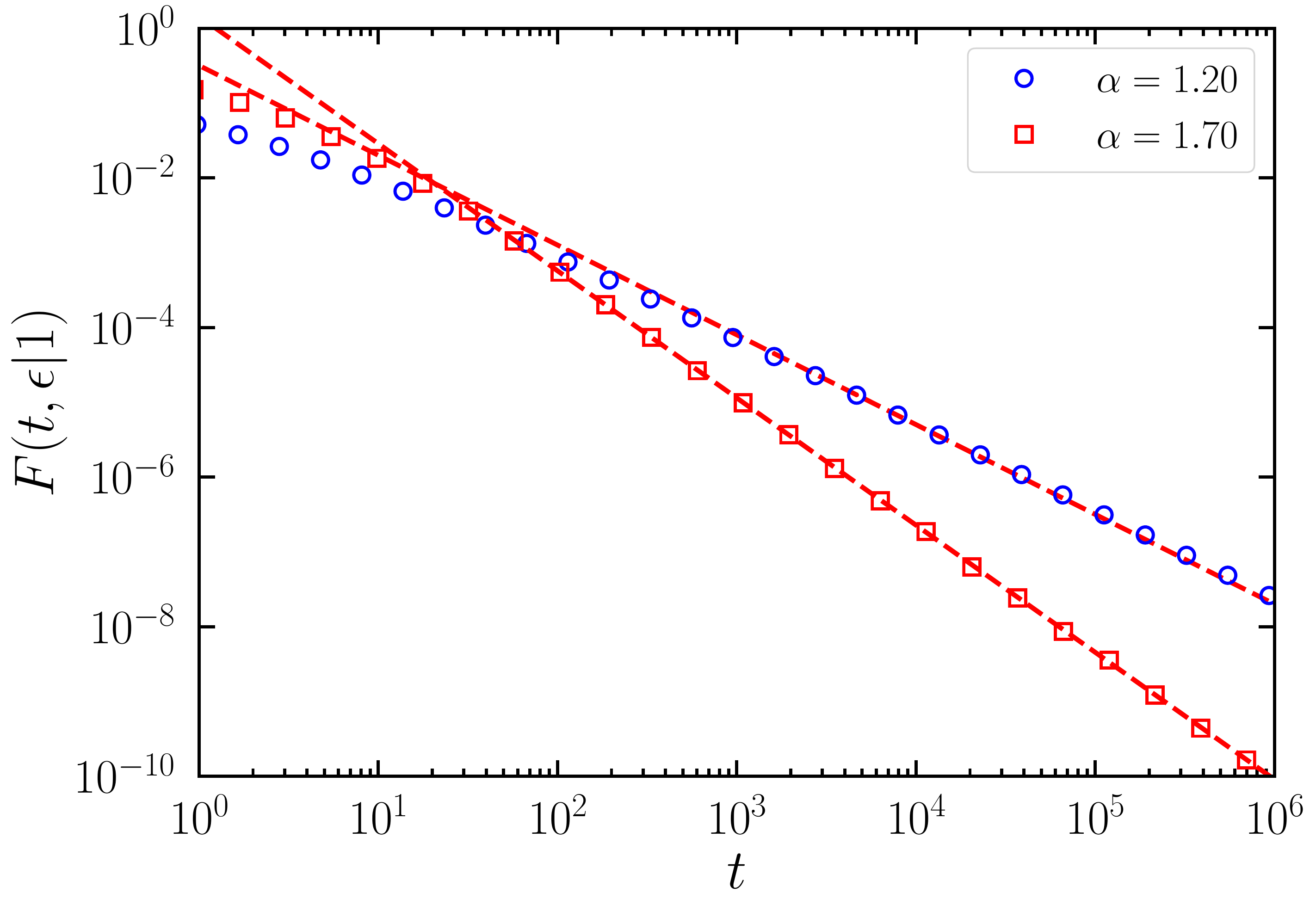}
  \caption{Distribution of the first passage time for a directed passive random
    walk on an infinitely aged NoPAD network, $F(t,c\vert c_0)$, with inter-event time distribution
    given by Eq.~\eqref{eq:d}, for different values of the exponent $\alpha$. We
    consider a bi-valued activity distribution with $c=\epsilon$ (with
    $\eta(\epsilon)=p$) or $c=1$ (with $\eta(1)=1-p$).  We plot the first
    passage time distribution to nodes with $c=\epsilon$, with the walker
    starting from nodes with $c_0=1$ (hollow symbols), along with the expected
    behavior $F(t,c\vert c_0)\sim t^{-\alpha}$ (dashed lines).  Number of
    walkers $W=10^6$, $p=0.5$, $\epsilon=0.1$.  Network size $N=10^5$.}
  \label{fig:FPT}
\end{figure}

For the first passage time distribution, Eq.~\eqref{eq:20} may be expanded in
the limit $s\ll 1$.  Inserting the expansion of the forward inter-event time
distribution Eq.~\eqref{eq:laplace_psi}, we obtain
\begin{equation}
\label{eq:fsc0}
  F(s,c\vert c_0) \simeq
  1-\left(c_0^{1-\alpha}+\frac{1}{\eta(c)}\sum_{c'\neq
      c}\eta(c')c'^{1-\alpha}\right)\Gamma_{2-\alpha}s^{\alpha-1}.
\end{equation}
In the time domain, this translates into a power-law behavior at large times,
$F(t,c\vert c_0)\sim t^{-\alpha}$.  This distribution lacks a first moment in
the regime $1<\alpha<2$, implying that the MFPT is infinite.  In
Fig.~\ref{fig:FPT} we perform numerical simulations to evaluate the first
passage time probability when the inter-event time distribution is power-law
with $1<\alpha<2$ and the activity is bi-valued, with $\eta(c)$ of the form
Eq.~\eqref{eq:bi-valued}.  For both values of $\alpha$ considered, one can
observe a power-law decay in the actual random walks (performed as described for
Fig.~\ref{fig:steadystate}), corresponding to the expected behavior in the
infinite network limit $F(t,c\vert c_0)\sim t^{-\alpha}$.  At the mean-field
level, the average time to reach a node with a given activity $c$ is equal to
the average number of independent trials required to land on a node with
activity $c$ (equal to $1/\eta(c)$), times the average waiting time spent on a
node.  Therefore, this time trivially diverges when the average waiting time is
infinite, as indicated here by a first passage time distribution lacking the
first moment.

\section{Non-Poissonian Activity-Driven networks with infinite average inter-event
  time}
\label{sec:infin-aver-inter}

We consider now an inter-event time distribution of the form Eq.~\eqref{eq:d}
with $0<\alpha<1$, which implies that the average time between two consecutive
activation of an agent with activity $c$ is infinite.  For such values of
$\alpha$, the dependency of the forward waiting time distribution on the aging
time cannot be eliminated even in the limit of strongly aged networks, so that
the use of the Laplace transform does not yield any substantial simplification.
Nevertheless, some insight may be obtained concerning the dynamics of the random
walk starting on a strongly aged network.

Let us recall the expression of the double Laplace transform of the forward
waiting time distribution,
namely~\cite{klafter_first_2011,aging,Barkai2003,godreche},
\begin{eqnarray}
  h_c(u,s) &=& \int_0^\infty dt  \int_0^\infty dt'  h_c(t',t) e^{-u t'}
  e^{-s t} \nonumber  \\
  &=& \frac{1}{1-\psi_c(u)}\frac{\psi_c(u)-\psi_c(s)}{s-u}.
  \label{eq:e}
\end{eqnarray}
Let us first consider the limit of strongly aged network and very large $t$,
with $ct \gg ct_a \gg 1$, corresponding to $s \ll u \ll 1$.  In this case, one
can expand
\begin{equation}
  \label{eq:39}
   h_c(u,s) \simeq \frac{u^\alpha - s^\alpha}{u^{1+\alpha}},
\end{equation}
which, upon inverse transformation, leads to
\begin{equation}
  \label{eq:40}
  h_c(t_a,t) \simeq t_a^{\alpha}\;\frac{\sin (\pi\alpha)}{\pi}\;t^{-\alpha-1}.
\end{equation}
On the other hand, in the limit of strong aging, $ct_a \gg 1$, but small
$t \ll t_a$, one can expand
$\psi_c(u) \sim 1 + \alpha \Gamma_{-\alpha} u^\alpha c^{-\alpha}$ in
Eq.~\eqref{eq:e} to obtain
\begin{equation}
  \label{eq:37}
  h_c(u,s) \simeq - \frac{1}{s} - \frac{1 - \psi_c(s)}{s \alpha \Gamma_{-\alpha}
  u^{\alpha} c^{-\alpha}},
\end{equation}
which, disregarding a constant term, leads to
\begin{equation}
  \label{eq:38}
  h_c(t_a,t) \simeq   c\,(c\,t_a)^{\alpha-1}\; \frac{\sin
    (\pi\alpha)}{\pi}\;\tilde{\psi}_c(t),
\end{equation}
where $\tilde{\psi}_c(t) = \int_t^\infty \psi_c(t')\, dt' = (ct+1)^{-\alpha}$.
The behavior of $\tilde{\psi}_c(t)$ can be approximated to
$\tilde{\psi}_c(t) \simeq 1$ if $c t \ll 1$, while for $c t \gg 1$, it holds
$\tilde{\psi}_c(t) \simeq (ct)^{-\alpha}$.  The behavior of $h_c(t_a,t)$ can
thus be summarized in the following three regimes:
\begin{equation}
  h_c(t_a,t) \simeq \left\{
    \begin{array}{ll}
     \frac{\sin (\pi\alpha)}{\pi}\; c\,(c\,t_a)^{\alpha-1} & \mathrm{for}  \; c\,t \ll
                                          1\\
       \frac{\sin (\pi\alpha)}{\pi}\;t_a^{\alpha-1}\;t^{-\alpha} &
                                                                 \mathrm{for}  \;\; 1 \ll c\,t \ll c\,t_a\\
      \frac{\sin (\pi\alpha)}{\pi}\;t_a^{\alpha}\;t^{-\alpha-1} & \mathrm{for}  \; t \gg t_a
    \end{array} \right. .
\label{eq:h_c}
\end{equation}

Interestingly, at large times, i.e. $ct \gg 1$, the forward waiting time
distribution is independent of $c$.  Besides, the tail of the distribution is
proportional to $t_a^{\alpha}\,t^{-\alpha-1}$, so that the probability that the
forward waiting time is greater than $t_a$ is constant and does not depend on
$t_a$.  This means that the interval $[t_a,+\infty[$ carries a constant
probability weight with respect to the other two terms, although its size
decreases when $t_a$ grows.  This, along with the fact that the total weight is
constant and equal to $1$ because $h_c$ is normalized, implies that the weight
carried in a time window $[0,t_0]$ tends to zero when $t_a$ tends to infinity.
In fact, one could argue that the weights calculated from Eq.~\eqref{eq:h_c} are
not exact because they neglect higher order corrections (in particular the
distribution in Eq.~\eqref{eq:h_c} is not normalized).  The reasoning is thus
true under the implicit assumption that the weights calculated from
Eq.~\eqref{eq:h_c} and carried in the intervals $[0,t_0]$ and $[t_0,+\infty[$
are proportional to their corresponding real weights, which is not guaranteed.

\begin{figure}[tbp]
  \centering \includegraphics[width=\columnwidth]{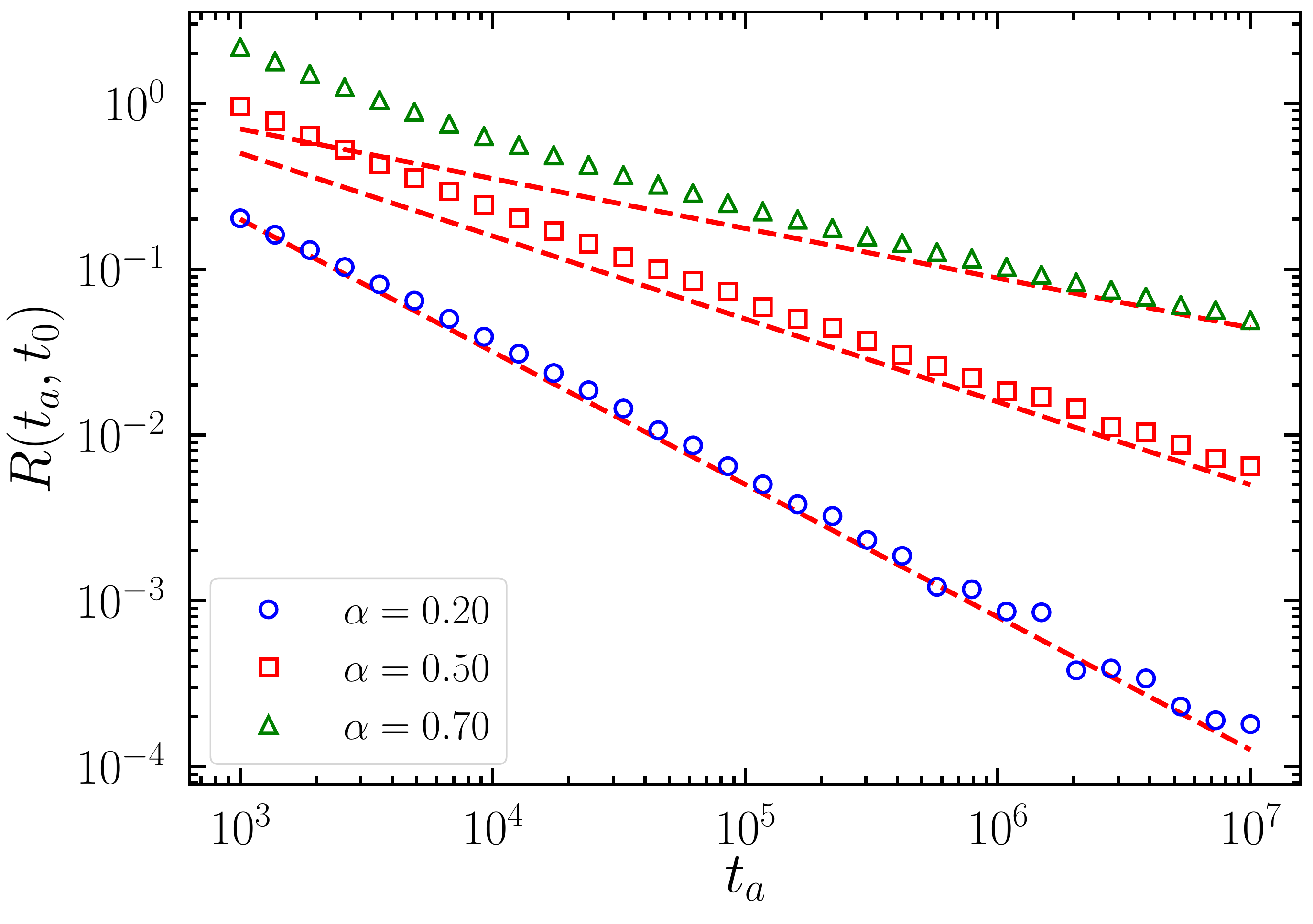}
  \caption{Ratios
    $R(t_a, t_0) =
    [\tilde{h}_c(t_a,0)-\tilde{h}_c(t_a,t_0)]/\,\tilde{h}_c(t_a,t_0)$ as a
    function of $t_a$, obtained from numerical simulations of a renewal process
    with an inter-event time distribution of the form Eq.~\eqref{eq:d} with
    different values of $\alpha$. The asymptotes $\alpha(t_0/t_a)^{1-\alpha}$
    calculated from Eq.~\eqref{eq:h_c} are plotted as dashed lines. Reference
    time $t_0=10^3$ and $c=1$.}
  \label{fig:ratio}
\end{figure}

In order to check these assumptions, on Fig.~\ref{fig:ratio} we compare the
ratio of the real weights
$R(t_a, t_0) \equiv
\left[\tilde{h}_c(t_a,0)-\tilde{h}_c(t_a,t_0)\right]/\,\tilde{h}_c(t_a,t_0)$,
evaluated from a numerical simulation of a renewal process with an inter-event
time distribution of the form Eq.~\eqref{eq:d}, and the ratio evaluated from
Eq.~\eqref{eq:h_c}, whose dominant order, with the conditions
$1\ll ct_0\ll ct_a$, is equal to $\alpha(t_0/t_a)^{1-\alpha}$.  We observe a
good agreement between the simulations and the analytical estimation, which
allows us to make the following reasoning: let us consider a NoPAD network with
inter-event time distribution given by Eq.~\eqref{eq:d} with $\alpha<1$, and an
arbitrary activity distribution $\eta(c)$ excluding zero-valued activities. Then
there exists a node with a minimum activity $c_{\mathrm{min}}>0$, and also a
time $t_0$, such that $c_{\mathrm{min}}\,t_0 \gg 1$. Then if the nodes are
synchronized at $t=-t_a$ with $t_a \gg t_0$ and we start an activated random
walk dynamics at time $t=0$, the probability that the time $t_1$ at which the
walkers escape from their first hosts is greater than $t_0$ is almost equal to
$1$. This holds a fortiori for all the following waiting times of the walker
occurring at times $t=t_2,t_3,...,t_k$ because $t_k$ is extracted from the
distribution $h_c(t_a+t_{k-1},\tau)$. Besides, the conditional probability that
$t_1=\tau$ given that $t_1 \geq t_0$ is independent of $c$ as we see from
Eq.~\eqref{eq:h_c}, which means that all the hops for all the walkers are
performed with waiting times that practically do not depend on the activity of
the hosts.

As a result, after its first jump, the probability that a walker is at a node of
activity $c$ is constant and equal to $\eta(c)$. In other words, if the initial
distribution of the walkers is $H(c)$, the probability $P(c,t)$ that the walker
is at a node with activity $c$ at time $t$ is equal to $\eta(c)$ if the walker
has escaped from its first host and $H(c)$ otherwise, i.e.
\begin{equation}
  P(c,t)\simeq H(c)\tilde{h}_c(t_a,t)+ \eta(c)(1-\tilde{h}_c(t_a,t)).
\label{eq:steady_state}
\end{equation}
In the limit of infinite $t$, $\tilde{h}_c(t_a,t)$ vanishes, and the steady
state of the walker is given by $P_\infty(c) = \eta(c)$. That is, in the large
time regime, the walker behaves as in a completely homogeneous network, in which
jumps were performed independently of the node activity. This result generalizes
the observation made in Ref.~\cite{2014arXiv1407.4582S}.
\begin{figure}[tbp]
  \centering \includegraphics[width=\columnwidth]{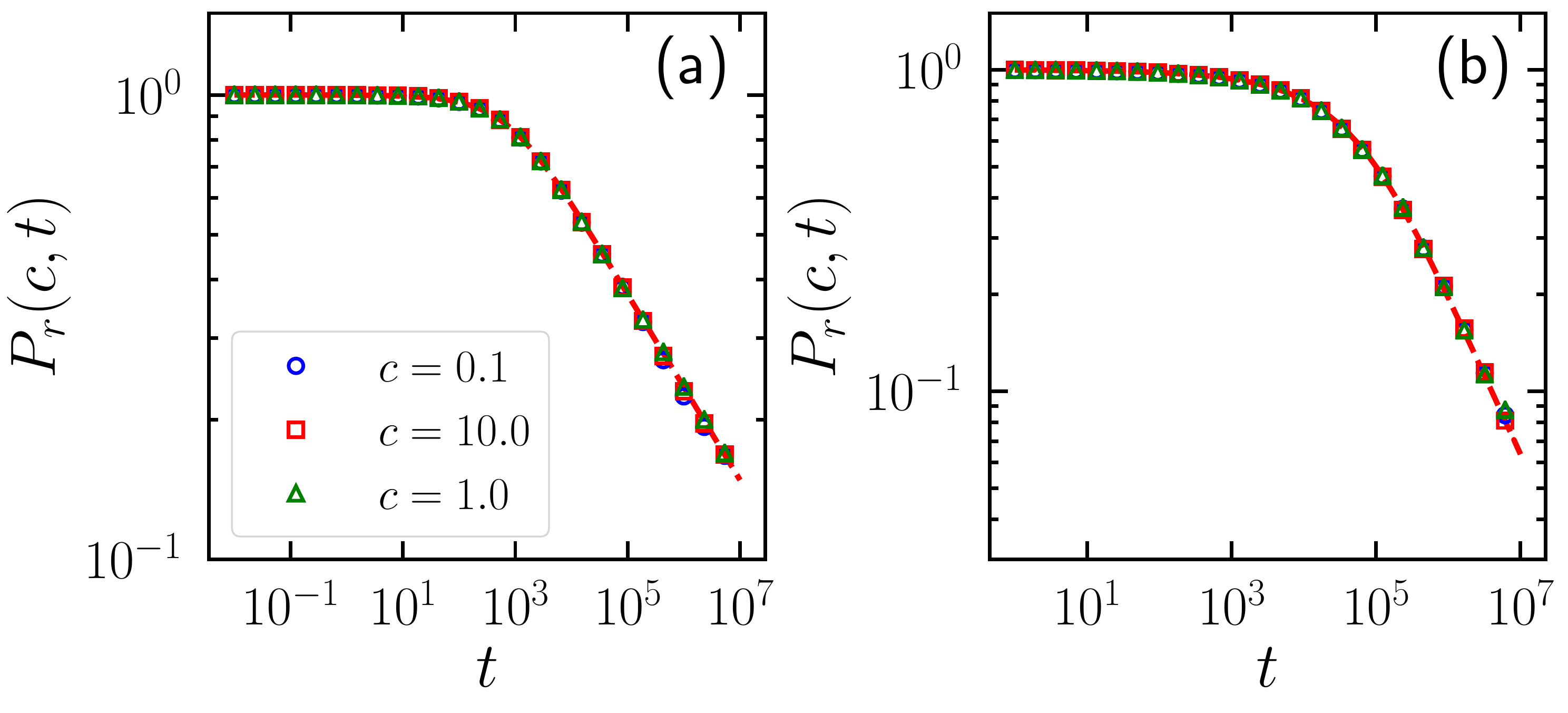}
  \caption{ Random walk dynamics on an aged NoPAD network with three-valued
    activity and infinite average waiting time. Activity values are $c=0.1$,
    $c=1$ and $c=10$, with $\eta(c)=1/3$. Initial distribution
    $H(c)=\delta_{c,10}$. We plot for the three values of $c$ the reduced
    occupation probability $P_r(c,t) = (P(c,t)-\eta(c))/(H(c)-\eta(c))$ as a
    function of time. (a): $\alpha=0.2$ and $t_a=10^3$. (b): $\alpha=0.5$ and
    $t_a=10^5$. The behavior predicted by Eq.~\eqref{eq:steady_state} is plotted
    in dashed lines. Network size $N=10^5$ and number of walkers $W=10^6$.}
  \label{fig:3states}
\end{figure}
In order to check the validity of the time dependence expressed in
Eq.~(\ref{eq:steady_state}), we have performed numerical simulations of the
activated random walk on a NoPAD network of size $N=10^5$ where the activity
takes three values, $c=0.1,1$ or $10$, each with probability $\eta(c) = 1/3$.
Walkers are initially hosted by nodes with activity equal to $c=10\;$,
i.e. $H(c)=\delta_{c,10}$.  Fig.~\ref{fig:3states} shows the reduced occupation
probability $P_r(c, t) = [P(c ,t)-\eta(c)]/[H(c)-\eta(c)]\;$ as a function of
the time $t$ for $\alpha=0.2$ and $t_a=10^3$, Fig.~\ref{fig:3states}(a), and for
$\alpha=0.5$ and $t_a=10^5$, Fig.~\ref{fig:3states}(b), along with their
expected value $\tilde{h}_{c}(t_a,t)$.  This last curve is  evaluated
from an independent numerical simulation of a renewal process, and is found to
be independent of the activity $c$.  We observe that the result stated in
Eq.~\eqref{eq:steady_state} perfectly match the numerical simulations in
networks of finite size.

\section{Conclusions}
\label{sec:conclusions}

In this paper we have explored the behavior of a passive node-centric random
walk unfolding on non-Markovian temporal networks generated by the NoPAD model,
which considers a power-law form $\psi_c(t) \sim (ct +1)^{-1-\alpha}$ of the
inter-event time distribution between consecutive activation events of nodes
with activity $c$. We have focused in particular on the behavior of the
occupation probability and first passage time distribution, in the case of a
very large aging time $t_a$, that is, when the time elapsed between the initial
synchronization of all nodes in the network and the start of the random walker
is very large.  The nature of the NoPAD model allows to simplify calculations in
the limit of infinite network size, in which every node in the path of the
walker is visited for the first time. In this approximation, we develop a
general theory for the walker dynamics, that can be analytically solved in
Laplace space if the inter-event time distribution of the nodes has a finite
first moment. In this case, in the limit $t_a \to \infty$, the waiting time of
the walker inside a node becomes independent of its arrival time, and a passive
random walk with inter-event time distribution $\phi_c(t) \sim t^{-1-\alpha}$,
with $\alpha>1$, behaves essentially as a active random walk with
$\phi_c(t) \sim t^{-\alpha}$, in which the internal clock of each node is reset
after the lading of the walker. Numerical simulations show that the actual
passive random walk process is very well described by our theory for a
sufficiently large network size.

If the inter-event time distribution lacks a first moment, which happens in the
case $\alpha<1$, our theory is not valid, since the waiting time inside a node
cannot be decoupled from the landing time. In the limit of very large $t_a$,
however, we develop arguments hinting that the random walker will ``feel'' a
network with homogeneous activity distribution, which implies that the
probability that the walker is at a node of activity $c$ is equal to $\eta(c)$
in the large time limit.  This result is straightforwardly extended to arbitrary
aging times $t_a$ (including non-aged networks $t_a=0$) because after a
transient regime of duration $t'$, the forward waiting time distribution
$h_c(t_a+t',\tau)$ will meet the conditions expressed in Eq.~\eqref{eq:h_c}, and
the system will be in the same situation as before, i.e. evolving as if the
network was homogeneous. This observation generalizes the results in
Ref.~\cite{2014arXiv1407.4582S} referred to networks with identical inter-event
time for all nodes.  Interestingly, this result is also recovered taking the
limit $\alpha \to 1$ in the equation describing the occupation probability in
the case of an inter-event time distribution with finite first moment, a fact
that provides additional evidence for its relevance.

\section*{Acknowledgments}

  This work was supported by the Spanish Government's MINECO, under project
  FIS2016-76830-C2-1-P. R.P.-S. acknowledges additional funding by ICREA
  Academia, funded by the \textit{Generalitat de Catalunya} regional
  authorities.

\end{document}